# High-index and low-loss topological insulators for mid-infrared nanophotonics

Sergey G. Menabde[1], Jacob T. Heiden[1], Vladimir A. Zenin[2], N. Asger Mortensen[3,4], and Min Seok Jang[1,*]

[1]School of Electrical Engineering, Korea Advanced Institute of Science and Technology, 34141 Daejeon, Korea
[2]Center for Nano Optics, University of Southern Denmark, DK-5230 Odense, Denmark
[3]POLIMA – Center for Polariton-driven Light-Matter Interactions, University of Southern Denmark, DK-5230 Odense, Denmark
[4]Danish Institute for Advanced Study, University of Southern Denmark, DK-5230 Odense, Denmark
*Corresponding author: jang.minseok@kaist.ac.kr

**Abstract**

Topological insulators generally have dielectric bulk and conductive surface states. Consequently, some of these materials have been shown to support polaritonic modes at visible and THz frequencies. At the same time, the optical properties of topological insulators in the mid-infrared (IR) remain poorly investigated. We employ near-field imaging to probe the mid-IR response from the exfoliated flakes of bismuth (Bi) / selenide (Se) / telluride (Te) / antimony (Sb) crystals with varying stoichiometry – $Bi_2Se_3$, $Bi_2Te_2Se$, and $Bi_{1.5}Sb_{0.5}Te_{1.7}Se_{1.3}$ – in pristine form as well as covered by thin flakes of hexagonal boron nitride (hBN). Contrary to theoretical expectations, all three materials exhibit a dielectric response with a high refractive index and with a loss below the experimental detection limit. Particularly, the near-field mapping of propagating phonon-polaritons in hBN demonstrates that these van der Waals crystals act as a practically lossless dielectric substrate with an ultra-high refractive index of up to 7.5 in $Bi_2Te_2Se$. Such a unique dielectric crystal would be of great advantage for numerous nanophotonic applications in the mid-IR.

# 1. Introduction

Strong spin-orbit coupling in topological insulators (TIs) leads to an electronic band inversion and the topological Dirac surface states in the bandgap [1,2]. These surface states are chiral and thus topologically protected from back-scattering by the time-reversal symmetry [3]. Therefore, charge carriers in surface states are free to move parallel to the surface and conduct current, while the insulating bulk remains dielectric. This particular property of TIs prompted significant interest from the photonics community as it promises exotic electrodynamic phenomena across a wide frequency spectrum.

The conductive surface state provides a condition for the manifestation of surface plasmons at optical frequencies in some TI materials [4]. For example, a particularly strong plasmonic response in visible and ultraviolet spectra has been reported in $Bi_{1.5}Sb_{0.5}Te_{1.8}Se_{1.2}$ (BSTS) [5,6]. Propagating surface plasmons in the visible spectrum have been directly observed with the scattering-type scanning near-field optical microscope (s-SNOM) [7] in $Bi_2Te_2Se$ (BTS) [8]. On the other hand, bulk $Bi_2Se_3$ (BS) is a polar dielectric and was predicted to support hyperbolic phonon-polaritons (HPP) in the THz regime [9] where its surface supports Dirac plasmons [10]. Later, the existence of the THz plasmon-phonon-polaritons in BS was confirmed by far-field experiments [11-13]. THz near-field imaging has been recently used to demonstrate the existence of the plasmon-phonon-polaritons in both BS [14,15] and BTS [14].

According to the theoretical models developed for these three TI materials (BS, BTS, and BSTS), all of them are expected to have a high-index low-loss dielectric bulk and a conductive surface at mid-infrared (IR) frequencies [4]. Other topological insulators such as $Bi_2Te_3$ [16,17] and Bi:Sb:Te family [18] are known to have a very high refractive index, but their bulk is lossy in the mid-IR [17].

Motivated by this, we aim to study the interaction between the highly-confined HPP in hexagonal boron nitride (hBN) and the topological surface states in BS, BTS, and BSTS of optimized chemical composition $Bi_{1.5}Sb_{0.5}Te_{1.7}Se_{1.3}$. To this end, we place thin (30–115 nm-thick) exfoliated flakes of hBN on top of the exfoliated TI crystals and obtain near-field images of propagating HPP using s-SNOM. Surprisingly, the measured HPP dispersion does not reveal any unambiguous signatures of the conductive surface states in any of the three materials. To confirm this observation, we analyze the complex near-field signal over the samples with hBN and across bare TI flakes on a silicon (Si) substrate. Again, the multifaceted near-field analysis does not exhibit any clear features of the conductive surface states in the TI crystals. At the same time, all experiments demonstrate the ultra-high refractive index and practically lossless nature of the three TI crystals in the tested mid-IR range (950 – 1600 cm$^{-1}$). Despite the possible existence of conductive surface states, our results suggest that in practice BS, BTS, and BSTS can be used as low-loss and ultra-high-index van der Waals dielectrics for numerous mid-IR applications[19,20].

## 2. Results

Optical images of exfoliated TI flakes with transferred hBN flakes on top are shown in Figure 1A. During the s-SNOM experiments, an atomic-force microscope (AFM) nanotip launches HPP waves that interfere with the same modes reflected at the hBN edge and form a near-field interference pattern with a period of $\lambda_{\text{HPP}}/2$, where $\lambda_{\text{HPP}}$ is the HPP wavelength [21-23]. Figure 1B shows an example of such an interference pattern imaged at 1500 cm$^{-1}$ in the 115 nm-thick hBN on 200 nm-thick BSTS flake. The near-field amplitude *s* is proportional to the amplitude of the out-of-plane component of the electric field, thus the interference pattern maps the electric field's amplitude of propagating polaritons (black curve in Fig. 1B). Assuming that the interference profile has a form of a damped harmonic oscillator, $s(x) = Ae^{-\gamma x}\cos(f_0 x)$, its Fourier spectrum can be accurately fitted with a Lorentzian-like profile which provides the polariton momentum $k = 2\pi/\lambda_{\text{HPP}} = 2\pi f_0$ and the decay constant *γ*, as illustrated in Fig. 1C.

Note that the *γ* extracted for the tip-launched polaritons with diverging wavefront (red curve in Fig. 1C) exceeds the actual decay constant of the HPP. In principle, the actual HPP damping can be estimated from the interference pattern formed by the edge-launched modes and the s-SNOM excitation beam [22] (blue curve in Fig. 1C). However, thin hBN edges have a small scattering cross-section and typically launch HPP less efficiently compared to the sharp metallized nanotip [21]. Therefore, in this work, we use the consistently strong signal from the tip-launched polaritons to extract HPP momentum and analyze the relative propagation loss.

For the analytical investigation, we employ the Drude model to describe the dielectric permittivity of the surface layer of finite thickness ~1.5 nm [4-6,8,24,25] and the Tauc–Lorentz model for the bulk BTS and BSTS [5,8]. Among the three materials, BSTS is predicted to have the highest surface conductivity comparable to that of a noble metal (Fig. 1D). At the same time, Ohmic loss in BS is predicted to be more than two orders of magnitude lower than in BSTS (Fig. 1B). Notably, the predicted permittivity of bulk BTS is as high as 57, while being practically dispersionless and lossless throughout mid-IR spectrum of interest (inset in Fig. 1D). The broadband dispersionless and lossless behaviors are qualitatively reflecting the principle of causality and Kramers–Kronig relations. The dielectric function of hBN is retrieved from Raman spectroscopy and near-field probing of a sample on a Si substrate. A detailed description of the employed models is provided in the Supplement.

The dispersion of eigenmodes in a continuous multilayer structure can be revealed by mapping the imaginary part of the reflection coefficient in momentum space, calculated by a generalized transfer-matrix method [26]. In addition, the transfer-matrix method provides the exact dispersion solution when formulated as an eigenvalue problem [22].

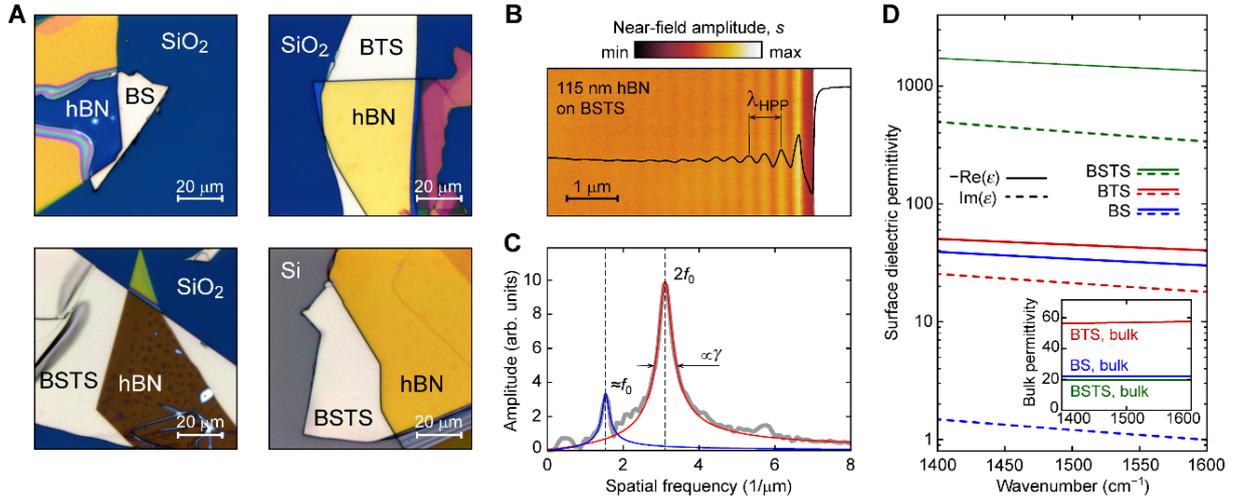

**Figure 1. A** Optical microscope images of hBN flakes placed of the TI flakes: 52, 113 and 30 nm-thick hBN on 275, 145, and 160 nm-thick BS, BTS, and BSTS, respectively, on Si/SiO$_2$ wafer; and 115 nm-thick hBN on 200 nm-thick BSTS on Si wafer. **B** Near-field amplitude at 1500 cm$^{-1}$ mapped over the hBN edge in the hBN/BSTS/Si sample. Profile of the near-field interference fringes (black curve) reveals the wavelength of the propagating phonon-polaritons. **C** Fourier spectrum of the interference profile in **B**, fitted by the Lorentzian-like spectrum of a damped oscillator for the nanotip-launched HPP (red) and the edge-launched HPP (blue). **D** Modelled 2D Drude permittivity of the topological surface layer in TI crystals at frequencies around the second Reststrahlen band of hBN. Inset: bulk dielectric permittivity according to the Tauc–Lorentz model for BTS and BSTS and the approximation for BS.

Figures 2A-D summarize the analytically calculated and measured dispersion of HPP in hBN for corresponding samples shown in Fig. 1A. The dispersion is calculated assuming a continuous multilayer structure for three cases: with the conductive surface layers (color map), without the surface layers (exact solution shown by red dash-dotted line), and for the suspended hBN (exact solution shown by grey dashed line). No additional fitting parameters are used to plot analytical dispersions in Fig. 2. The modelled permittivity of surface layers in BS and BTS

appears to be too low to cause any significant perturbation of the HPP dispersion (Fig. 2A, B). However, much more conductive BSTS surfaces are expected to support plasmonic modes that can hybridize with the HPP and significantly perturb the dispersion (color map in Fig. 2C, D and Fig. S3 in the Supporting Information). Therefore, we studied three samples with 30, 86, and 115 nm-thick hBN on BSTS to probe different regions of the momentum space (dispersion for 86 nm-thik hBN is shown in Supporting Information Fig. S4).

As expected, the HPP in all samples are significantly more confined compared to the free-standing hBN due to the high-index TI substrates. To our surprise however, experimental results in every case are explained by the analytical dispersion that neglects possible contributions from the theoretically anticipated conductive surface layers. Further discussion is focused on investigating this unexpected observation.

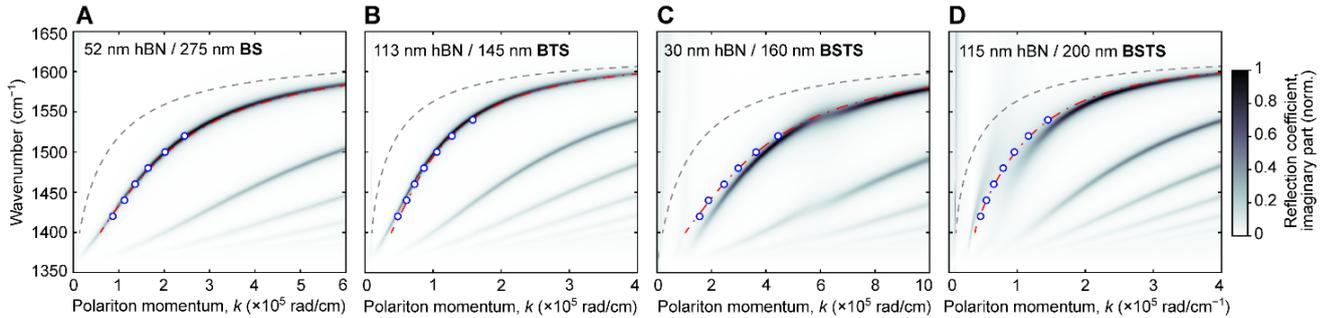

**Figure 2.** Dispersion of HPP in the samples shown in Fig. 1: calculated assuming TI crystals with conductive surface layers (color map), without the surface layers (red dash-dotted), and experimentally measured (data points). **A-C** Samples on Si/SiO$_2$ wafer. **D** Sample on Si wafer. Also shown is the HPP dispersion in a free-standing hBN of corresponding thickness (grey dashed).

To commence, we note that any conductive surface states are expected to cause a slightly higher propagation loss of HPP compared to hBN on a low-loss substrate, thus also potentially providing a means for its experimental detection. As a figure of merit (FOM) for the HPP loss, we use ratio $k/2\pi\gamma$ that gives a normalized propagation length in units of polariton wavelengths. Since TI materials are van der Waals crystals, we selected flakes with flat surfaces (by using dark-field microscopy for inspection) to minimize the scattering-mediated loss. Figure 3 shows the HPP FOM in hBN on TI crystals and on a Si wafer which is practically lossless and dispersionless at the frequencies of interest [27]. In notable agreement, the FOM on TI flakes is very similar to that on the Si wafer, indicating the absence of additional loss.

Summarizing the near-field study of the propagating HPP in hBN at 1420–1540 cm$^{-1}$, we conclude that the investigated TI materials behave as low-loss dielectrics with the refractive index correctly predicted by the Tauc–Lorentz model. At the same time, any interaction between the conductive surface states and HPP remains undetected even when polaritons hybridization is expected in BSTS.

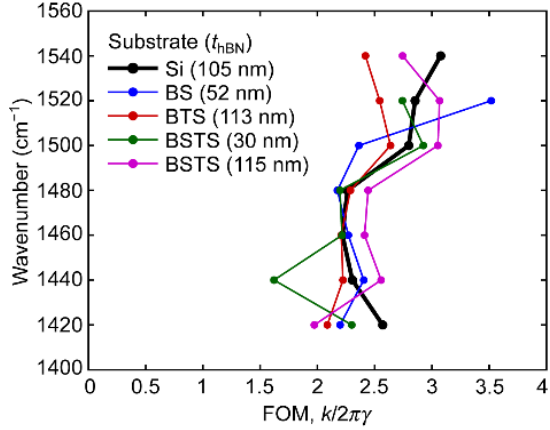

**Figure 3.** Figure of merit for the HPP propagation loss in the samples shown in Fig. 1A (colors) and that in the hBN on a silicon substrate (black).

To further investigate our intriguing findings, we analyze the complex-valued s-SNOM signal $\sigma = se^{i\varphi}$ over different areas of the samples far away from edges, which carries information about the local material properties under the nanotip. The amplitude $s$ is proportional to the near-field reflectivity of the material, and its phase $\varphi$ has been associated with the loss in the sample [28]. At the same time, the background signal in s-SNOM (i.e. not from the near-field interaction) is suppressed by demodulation of $\sigma$ at higher-order harmonics of the AFM tapping amplitude [29]. Due to the highly nonlinear dependency of $\sigma$ on the tip-sample distance, the higher-order harmonics $\sigma_{n>2}$ probe the near-field within a smaller volume under the tip [30-32], as schematically shown in Fig. 4A. Hence their phase $\varphi_{n>2}$ corresponds to the electromagnetic loss in a different volume, depending on $n$. The phase ambiguity caused by the random but constant term from the reference arm of the SNOM's interferometer can be eliminated by considering the phase difference $\varphi_{n+1} - \varphi_n$. Particularly, it has been demonstrated that $\varphi_{n+1} - \varphi_n > 0$ in the presence of a thin layer of lossy material on top of the sample, and $\varphi_{n+1} - \varphi_n = 0$ when the sample is lossless [30-32]. Therefore, the presence of the Drude layer (black dashed line in Fig. 4a) would lead to $\varphi_{n+1} - \varphi_n > 0$ even if the bulk would be lossless. Here, we apply this technique in an attempt to detect the elusive surface state in BSTS.

Figure 4B shows the map of $\varphi_4 - \varphi_3$ over the hBN edge on BSTS at 1500 cm$^{-1}$ in the two samples with hBN thickness $t_{hBN}$ = 30 and 115 nm. We note that the thickness of BSTS in both samples is larger than 150 nm, which exceeds the near-field probing depth for $n > 2$ [30,33], thus isolating the signal from the different substrates under BSTS. Corresponding spatial profiles of $\varphi_4 - \varphi_3$ are shown in Fig. 4C. First, $\varphi_4 - \varphi_3 \approx 0.4°$ and $–0.5°$ over the BSTS in the sample with thinner and thicker hBN, respectively. Second, $\varphi_4 - \varphi_3 > 0$ far from the hBN edge (where the HPP interference does not contribute to near-field signal), and is larger in the sample with thinner hBN as expected from a thinner absorbing layer [30]. This concludes that the hBN is responsible for the near-field loss probed by $\varphi_4$ and $\varphi_3$, while BSTS is practically lossless.

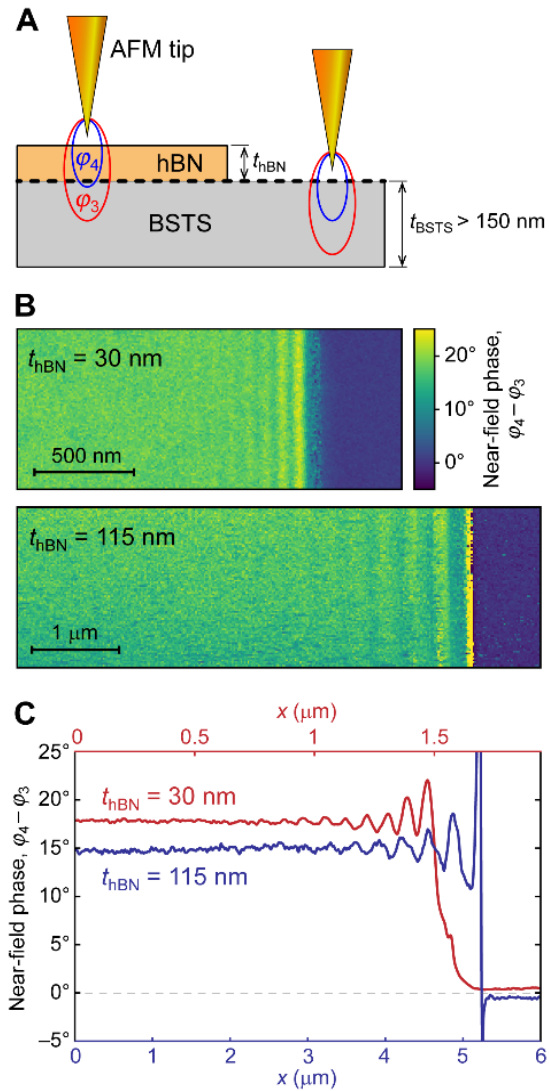

**Figure 4. A** Schematic demonstration of how the higher-order harmonics of the near-field signal probe the smaller volume under the nanotip in the sample. This leads to $\varphi_{n+1} - \varphi_n > 0$ over the lossy sample areas, including the hypothetical case when only the surface state is lossy (dashed black line). **B** Maps of $\varphi_4 - \varphi_3$ over the hBN edge on BSTS in the two different samples mapped at 1500 cm$^{-1}$. **C** Spatial profiles of $\varphi_4 - \varphi_3$ averaged across the scan areas shown in **B**.

Finally, we conduct near-field contrast analysis for all three (bare) TI materials on Si substrate at two frequencies 950 cm$^{-1}$ and 1600 cm$^{-1}$. Since Si can be considered practically lossless across the mid-IR spectrum, the imaginary part of the near-field contrast $\eta = \sigma/\sigma_{Si}$ over the TI flakes would reveal the presence of any loss.

Figure 5 summarizes the near-field contrast of the three TI flakes at 950 cm$^{-1}$ (data at 1600 cm$^{-1}$ is provided in the Supporting Information Fig. S5). The real part of the near-field contrast $\text{Re}(\eta) = s/s_{Si}$ (Fig. 5A) is proportional to the tip polarizability, which is a function of dielectric permittivity [30]. In agreement with the Tauc–Lorentz model, $s/s_{Si}$ is significantly higher over BTS, resulting from the very high bulk permittivity ($\varepsilon_{BTS} \approx 57$), while $s_{BS}/s_{Si} \approx 1$ indicates that the permittivity of BS is approximately equal to that of Si ($\varepsilon_{Si} = 12.04 + 5i \times 10^{-4}$ at 950 cm$^{-1}$ [27]). In principle, near-field contrast can be used to extract the dielectric function of an unknown material by probing the samples of different thicknesses [33-35]. However, this task is prohibitively complex and is out of the scope of this work. The normalized phase (Fig. 5B) does not exhibit any clear contrast between the TI flakes and Si, and in general, it is close to zero over all samples, which once again indicates the near-zero loss.

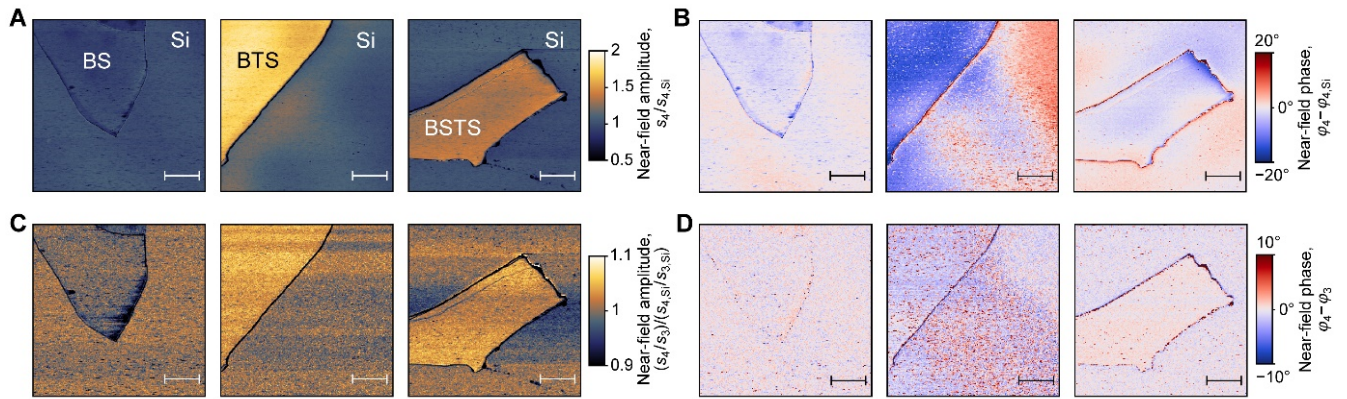

**Figure 5.** Spatial distributions of **A** amplitude and **B** phase of the near-field contrast calculated by normalizing the near-field signal by its average over Si. **C** Near-field amplitude ratio between the 4$^{th}$ and 3$^{rd}$ demodulation harmonics normalized by that over Si, thus eliminating the effects of non-uniform tip illumination. **D** Phase difference between the 4$^{th}$ and 3$^{rd}$ demodulation harmonics of the near-field signal across the whole scan areas. All scale bars are 2 μm. Flakes thickness: BS 31 nm, BTS 188 nm, BSTS 91 nm.

The very high permittivity of BTS and BSTS causes the non-uniform illumination of the tip as it scans over the area [31], leading to the fluctuation of the near-field signal in Fig. 5A, B. Since the non-uniform illumination proportionally contributes to all demodulation harmonics of the output signal, it can be eliminated by mapping the ratio $\sigma_{n+1}/\sigma_n = (s_{n+1}/s_n)e^{i(\varphi_{n+1}-\varphi_n)}$. The normalized amplitude ratio $(s_4/s_3)/(s_{4,\text{Si}}/s_{3,\text{Si}})$ is shown in Fig. 5C, revealing the flakes edges without the signal fluctuations. Most importantly, as we discussed earlier, $\varphi_{n+1}$ and $\varphi_n$ reflect the near-field loss within different volumes under the tip. Thus, in a lossless material, we can expect $\varphi_{n+1} - \varphi_n = 0$. Figure 5D maps the phase difference $\varphi_4 - \varphi_3$ with much smaller signal fluctuations compared to $\varphi_4 - \varphi_{4,\text{Si}}$ (Fig. 5B). Yet again, $\varphi_4 - \varphi_3 \approx 0$ across all three TI flakes, thus confirming our earlier observations.

At the same time, it has been demonstrated [36] that the mid-IR near-field phase contrast between a dielectric substrate and a 2D conductive layer should be detectable when the 2D carrier concentration exceeds ~1×10$^{12}$ cm$^{-2}$, even if carrier mobility is as low as ~1 cm$^2$V$^{-1}$s$^{-1}$. However, this carrier concentration is approximately two orders of magnitude smaller than theoretically predicted for any of the three TI crystals, and about an order of magnitude smaller than measured experimentally [4].

## 3. Conclusion

In conclusion, we studied the mid-IR electrodynamic response of three van der Waals crystals – BS, BTS, and BSTS – by several near-field probing techniques. Surprisingly, the conductive surface state was not detected by any experiment in all three TI materials. We speculate that such lossless mid-IR response can be caused by unusually low carrier concentration in the surface states, or possibly by a not-Drude-like behavior of the topological surface conductivity in the poorly investigated mid-IR regime. This puzzle demands an investigation but is out of the scope of this work.

Nevertheless, our results have significant implications for mid-IR nanophotonics: we show that these van der Waals crystals have ultra-low loss (comparable to silicon) and ultra-high refractive index (as much as 7.5 in BTS) across a wide mid-IR frequency range from 950 cm$^{-1}$ to 1600 cm$^{-1}$. For example, as demonstrated by our experiments, exfoliated TI flakes provide a low-loss substrate for highly-confined phonon-polaritons, which generally have a much higher quality factor than plasmons. Thus, our results open new avenues for mid-IR nanophotonic applications and hint at yet-to-be-explored physical properties of certain topological insulators.

## 4. Methods

*Sample Preparation:* The bulk BS, BTS, and BSTS crystals grown by the Bridgman method were commercially purchased from 2D Semiconductors USA. Mechanically exfoliated (with a low adhesion Nitto SPV224 tape) thin flakes were transferred on top of the wafer using a polydimethylsiloxane (PDMS) stamp and a manual transfer stage. Large-size bulk hBN crystal grown by the epitaxial solidification technique was also purchased from 2D Semiconductors USA, and then exfoliated and transferred in the same way as the TI flakes.

*Sample Characterization:* s-SNOM measurements were performed using the neaSNOM from attocube systems AG (formerly Neaspec) coupled with the tunable QCL (MIRcat, Daylight Solutions). The Pt-coated AFM tips (ARROWNCPt, Nano World) had a typical tapping frequency $\Omega$ around 270 kHz, and the used taping amplitude was 60–70 nm in a tapping mode. The background-free interferometric signal [29] demodulated at the third and the fourth harmonics ($3\Omega$ and $4\Omega$) was used to generate the near-field images and analyze the near-field data. During the near-field imaging, the topography of the sample was also recorded, which was used to measure the thickness of the flakes.